\documentclass[preprint,12pt]{elsarticle}
\biboptions{sort&compress}
\usepackage{array}
\usepackage{longtable,booktabs}
\usepackage{amsmath,bbold,amssymb,epsfig,bm,mathrsfs}
\usepackage{feynmp,color,slashed,nicefrac,exscale,multirow,graphicx}
\usepackage{epstopdf}
\usepackage{longtable}
\usepackage{multirow}

\usepackage[dvipdfm,bookmarks=true,colorlinks,%
            citecolor=blue,linkcolor=blue,hypertex, %
            breaklinks=true]{hyperref}

\journal{Nuclear Physics A }

\begin{document}

\begin{frontmatter}
\hyphenpenalty=6000
\tolerance=1000

\title{Systematic investigation of the high-spin structures
in the odd-odd nuclei $^{166, 168, 170, 172}$Re by a particle-number
conserving method}

\author{Zhen-Hua Zhang\fnref{contact}}
  \ead{zhzhang@ncepu.edu.cn}
  \cortext[cor1]{Corresponding author.}

\author{Shuo-Yi Liu}

\address{Mathematics and Physics Department,
              North China Electric Power University, Beijing 102206, China}

\date{\today}

\begin{abstract}
The recently observed two and four-quasiparticle high-spin rotational bands
in the odd-odd nuclei $^{166, 168, 170, 172}$Re are investigated
using the cranked shell model with pairing correlations treated
by a particle-number conserving method.
The experimental moments of inertia and alignments can be reproduced
well by the present calculation if appropriate bandhead spins and
configurations are assigned for these bands, which in turn confirms
their spin and configuration assignments.
It is found that the bandhead spins of those two rotational
bands observed in $^{166}$Re~[Li {\it et al.}, Phys. Rev. C 92 014310 (2015)]
should be both increased by $2\hbar$ to get in consistent with the
systematics of the experimental and calculated moments of inertia for the
same configurations in $^{168, 170, 172}$Re.
The variations of the backbendings/upbendings with increasing neutron number
in these nuclei are investigated.
The level crossing mechanism is well understood by
analysing the variations of the occupation probabilities of the
single-particle states close to the Fermi surface and their
contributions to the angular momentum alignment with rotational frequency.
In addition, the influence of the deformation driving effects of the proton
$1/2^-[541]$ ($h_{9/2}$) orbtial on the level crossing
in $^{172}$Re is also discussed.
\end{abstract}


\end{frontmatter}


\section{Introduction}{\label{Sec:Introduction}}

The high-spin structures of the rare-earth nuclei have drawn
a lot of attentions owing to the existence of various exotic excitation modes,
such as backbending~\cite{Johnson1971_PLB34-605, Lee1977_PRL38-1454},
signature inversion~\cite{Bengtsson1984_NPA415-189},
band termination~\cite{Bengtsson1983_PST5-165, Afanasjev1999_PR322-1},
superdeformation~\cite{Twin1986_PRL57-811},
wobbling mode~\cite{Odegard2001_PRL86-5866}, etc.

Compared to even-even and odd-$A$ nuclei, the doubly-odd nuclei
often show a broader variety of nuclear structure phenomena,
and they are more challenging to study due to the complexity associated
with couplings from both valence quasiproton and quasineutron.
Consequently, only limited information is available for odd-odd nuclei
throughout the whole nuclear chart.
Usually, it is challenging to make spin-parity and configuration
assignments for the states in odd-odd nuclei,
since there always exist a high density of low-lying states,
which leads to significant level mixing.

Recently, a considerable amount of two- and four-quasiparticle (4-qp)
rotational bands in doubly-odd nuclei $^{166, 168, 170, 172}$Re~\cite{Li2015_PRC92-014310,
Hartley2016_PRC94-054329, Hartley2013_PRC87-024315, Hartley2014_PRC90-017301}
have been observed experimentally.
These neutron-deficient Re ($Z=75$) isotopes are characterized by
fairly small quadrupole deformations about $\beta_2 \sim 0.2$.
Their proton and neutron Fermi surfaces lie close to the
$\pi h_{11/2}$ and $\nu i_{13/2}$ sub-shells, respectively,
which provides a good opportunity to study the dependence of
level crossings and angular momentum alignments on
the occupation of specific single-particle orbitals.
Moreover, these data provide an excellent testing ground for various
nuclear cranking models, e.g., the cranked Nilsson-Strutinsky
method~\cite{Andersson1976_NPA268-205}, the cranking Hartree-Fock-Bogoliubov (HFB)
model with Nilsson~\cite{Bengtsson1979_NPA327-139}
and Woods-Saxon potentials~\cite{Nazarewicz1985_NPA435-397, Cwiok1987_CPC46-379},
the projected shell model~\cite{Hara1995_IJMPE4-637},
the tilted axis cranking model~\cite{Frauendorf2001_RMP73-463},
the cranked non-relativistic~\cite{Terasaki1995_NPA593-1,
Egido1993_PRL70-2876, Afanasjev2000_PRC62-054306},
and relativistic mean-field models~\cite{Afanasjev1996_NPA608-107,
Afanasjev2000_PRC62-054306}, etc.

In Ref.~\cite{Li2015_PRC92-014310}, two rotational bands with the
configurations assigned as $\pi h_{11/2} \otimes \nu i_{13/2}$
and $\pi h_{11/2} \otimes \nu [f_{7/2} / h_{9/2}]$ have been observed in
the lightest odd-odd rhenium nucleus $^{166}$Re.
Their bandhead spins are assigned as $8^-$ and $6^+$, respectively.
Later on, two rotational bands with the same configurations have been
identified in $^{168}$Re by D. J. Hartley et al~\cite{Hartley2016_PRC94-054329}.
According to the energy level systematics and the additivity of alignment,
they suggested that the bandhead spins of the two bands observed in $^{166}$Re
should be $10^-$ and $7^+$, respectively.
Similar band structures have also been observed in $^{170}$Re~\cite{Hartley2013_PRC87-024315}.
While in $^{172}$Re, except $\pi h_{11/2} \otimes \nu i_{13/2}$,
several multi-qp configurations including the proton $h_{9/2}$ orbital
have been observed in Ref.~\cite{Hartley2014_PRC90-017301}.
It is well known that $h_{9/2}$ orbital has very strong deformation driving effects,
which makes the level crossings and alignments in $^{172}$Re extremely complicated.
It should be noted that several configuration assignments in these odd-odd Re isotopes
are proposed based on alignment properties and observed band crossings.
Therefore, it is important to determine their bandhead spins and configurations,
as well as the level crossing mechanism.

In this paper, the cranked shell model (CSM) with pairing correlations
treated by a particle-number conversing (PNC) method~\cite{Zeng1983_NPA405-1, Zeng1994_PRC50-1388}
will be used to investigate the recently observed 2- and 4-qp
high-spin rotational bands in $^{166, 168, 170, 172}$Re.
In the PNC method, the pairing Hamiltonian is diagonalized directly in a
properly truncated Fock space~\cite{Wu1989_PRC39-666, Molique1997_PRC56-1795}.
Therefore, the particle-number is strictly conserved and the Pauli blocking effects are
taken into account exactly, which makes it especially suitable for the investigation of the
high-spin states in doubly-odd nuclei~\cite{He2005_NPA760-263, Li2013_ChinPhysC37-014101,
Zhang2016_SciChinaPMA59-672012, Zhang2019_NPA981-107, Liu2019_PRC100-064307}.
Note that the PNC scheme has also been transplanted in
the total-Routhian-surface method~\cite{Fu2013_PRC87-044319},
the cranking Skyrme-Hartree-Fock method~\cite{Liang2015_PRC92-064325}, and the cranking
covariant density functional theory~\cite{Shi2018_PRC97-034317, Xiong2020_PRC101-054305}
for investigating the high-spin states and the high-$K$ isomers.
The PNC-CSM with octupole deformation has also been developed~\cite{He2020_PRC102-064328}.
Recently, a detailed comparison of different mean fields and treatments of
pairing correlations in the description of the rotational excitations have been
performed~\cite{Zhang2020_PRC101-054303}.
Similar approaches with exactly conserved particle number when treating the pairing correlations
can be found in Refs.~\cite{Richardson1964_NP52-221, Pan1998_PLB422-1, Volya2001_PLB509-37,
Pillet2002_NPA697-141, Jia2013_PRC88-044303, Jia2013_PRC88-064321, Chen2014_PRC89-014321}.

This paper is organized as follows.
In Sec.~\ref{Sec:PNC-CSM}, a brief introduction to PNC-CSM is presented.
Sec.~\ref{Sec:Num} shows the numerical details of the calculation for Re isotopes.
The comparison of experimental and calculated moments of inertia (MOIs) and alignments
for the rotational bands in $^{166,168,170,172}$Re are shown in Sec.~\ref{Sec:Results}.
The level crossing mechanism in these bands is also analyzed.
A brief summary is given in Sec.~\ref{Sec:Summary}.

\section{Theoretical framework}{\label{Sec:PNC-CSM}}

The CSM Hamiltonian with pairing correlations is written as
\begin{eqnarray}
 H_\mathrm{CSM}
 & = &
 H_0 + H_\mathrm{P}
 = H_{\rm Nil}-\omega J_x + H_\mathrm{P}
 \ ,
 \label{eq:H_CSM}
\end{eqnarray}
where $H_{\rm Nil}$ is the Nilsson Hamiltonian~\cite{Nilsson1969_NPA131-1},
and $-\omega J_x$ is the Coriolis interaction with
the rotational frequency $\omega$ about the $x$ axis.
$H_{\rm P}$ is the monopole pairing interaction,
\begin{eqnarray}
 H_\mathrm{P}
 & = &
 -G\sum_{\xi\eta}a^{\dagger}_{\xi}a^{\dagger}_{\bar{\xi}}a_{\bar{\eta}}a_{\eta}
 \ ,
 \label{eq:H_p}
\end{eqnarray}
where $\bar{\xi}$ $(\bar{\eta})$ denotes the time-reversal
state of a Nilsson state $\xi$ ($\eta$),
and $G$ is the effective monopole pairing strength.

Instead of the single-particle level truncation in traditional
shell-model calculations, a cranked many-particle configuration
(CMPC, the eigenstates of the one-body operator $H_0$)
truncation (Fock space truncation) is adopted, which is crucial
to make the PNC-CSM calculations both workable and sufficiently
accurate~\cite{Molique1997_PRC56-1795, Wu1989_PRC39-666}.
Usually the CMPC space with about 1000 dimension is enough for the investigation
of the yrast and low-lying excited states in the rare-earth nuclei.
In contrast to the conventional Bardeen-Cooper-Schrieffer or
Hartree-Fock-Bogoliubov approaches, without introducing quasiparticle transformation,
the pairing Hamiltonian is diagonalized directly in PNC-CSM.
Therefore, the particle-number is conserved from beginning to the end
and the Pauli blocking effects are treated exactly.
Especially, the odd-$A$ and odd-odd nuclei are treated
at the same footing as the even-even ones in PNC-CSM,
which makes the calculations much easier and more reliable.

By diagonalizing the many-body Hamiltonian in a sufficiently large CMPC space,
the eigenstates of $H_{\rm CSM}$ can be obtained as
\begin{equation}
 |\Psi\rangle = \sum_{i} C_i \left| i \right\rangle
 \qquad (C_i \; \textrm{real}) \ ,
\end{equation}
where $| i \rangle$ is a CMPC,
and $C_i$ is the corresponding expansion coefficient.

The angular momentum alignment for the state $| \Psi \rangle$ is
\begin{equation}
\langle \Psi | J_x | \Psi \rangle = \sum_i C_i^2 \langle i | J_x | i
\rangle + 2\sum_{i<j}C_i C_j \langle i | J_x | j \rangle \ ,
\end{equation}
and the kinematic MOI is
\begin{equation}
J^{(1)}=\frac{1}{\omega} \langle\Psi | J_x | \Psi \rangle \ .
\end{equation}
Because $J_x$ is a one-body operator, the matrix element $\langle i | J_x | j \rangle$
($i\neq j$) may not vanish only when two CMPCs $|i\rangle$ and $|j\rangle$ differ by
one particle occupation~\cite{Zeng1994_PRC50-1388}.
After a certain permutation of creation operators,
$|i\rangle$ and $|j\rangle$ can be recast into
\begin{equation}
 |i\rangle=(-1)^{M_{i\mu}}|\mu\cdots \rangle \ , \qquad
|j\rangle=(-1)^{M_{j\nu}}|\nu\cdots \rangle \ ,
\end{equation}
where $\mu$ and $\nu$ denotes two different single-particle states,
and $(-1)^{M_{i\mu}}=\pm1$, $(-1)^{M_{j\nu}}=\pm1$ according to
whether the permutation is even or odd.
Therefore, the angular momentum alignment of $|\Psi\rangle$ can be written as
\begin{equation}
 \langle \Psi | J_x | \Psi \rangle = \sum_{\mu} j_x(\mu) + \sum_{\mu<\nu} j_x(\mu\nu)
 \ ,
 \label{eq:jx}
\end{equation}
where the diagonal contribution $j_x(\mu)$ and the
off-diagonal contribution $j_x(\mu\nu)$ reads
\begin{eqnarray}
j_x(\mu)&=&\langle\mu|j_{x}|\mu\rangle n_{\mu} \ , \nonumber
\\
j_x(\mu\nu)&=&2\langle\mu|j_{x}|\nu\rangle\sum_{i<j}(-1)^{M_{i\mu}+M_{j\nu}}C_{i}C_{j}
  \quad  (\mu\neq\nu) \ ,
 \label{eq:jxorb}
\end{eqnarray}
and
\begin{equation}
n_{\mu}=\sum_{i}|C_{i}|^{2}P_{i\mu} \ ,
\end{equation}
is the occupation probability of the cranked Nilsson orbital $|\mu\rangle$,
$P_{i\mu}=1$ if $|\mu\rangle$ is occupied in the CMPC $|i\rangle$,
and $P_{i\mu}=0$ otherwise.

The experimental kinematic MOIs are extracted separately
for each signature sequence within a rotational band ($\alpha = I$ mod 2) using
\begin{eqnarray}\label{eq:exp-moi}
\frac{J^{(1)}(I)}{\hbar^2}&=&\frac{2I+1}{E_\gamma(I+1\rightarrow I-1)} \ , \nonumber \\
\hbar\omega(I)&=&\frac{E_{\gamma}(I+1\rightarrow I-1)}{I_x(I+1)-I_x(I-1)} \ ,
\end{eqnarray}
where $I_x(I)=\sqrt{(I+1/2)^2-K^2}$, $K$ is the projection of the total
angular momentum onto the symmetry axis.

\section{Numerical details}{\label{Sec:Num}}

\begin{table}[h]
 \centering
 \caption{\label{tab:def}
Deformation parameters ($\varepsilon_2$, $\varepsilon_4$) for Re
isotopes in the present PNC-CSM calculation,
which are taken from Ref.~\cite{Bengtsson1986_ADNDT35-15}
as an average of the adjacent even-even nuclei.}
\begin{tabular*}{0.7\columnwidth}{c@{\extracolsep{\fill}}ccccc}
\hline
\hline
~               & $^{166}$Re & $^{168}$Re & $^{170}$Re & $^{172}$Re  \\
\hline
$\varepsilon_2$ & 0.125      & 0.168      & 0.196      & 0.213       \\
$\varepsilon_4$ & -0.001     & 0.000      & 0.002      & 0.008       \\
\hline
\hline
\end{tabular*}
\end{table}

\begin{table}[h]
 \centering
 \caption{\label{tab:ku}
The Nilsson parameters ($\kappa$, $\mu$) for Re isotopes,
which are taken from Ref.~\cite{Bengtsson1985_NPA436-14}
with a slight modification of neutron $\mu_6$ from 0.34 to 0.28.}
\begin{tabular*}{0.7\columnwidth}{c@{\extracolsep{\fill}}ccccc}
\hline
\hline
~                       & $N$       & 4      & 5     & 6      \\
\hline
\multirow{2}*{Protons}  & $\kappa$  & 0.065  & 0.060 & ~      \\
                        & $\mu$     & 0.57   & 0.65  & ~      \\
\hline
\multirow{2}*{Neutrons} & $\kappa$  & ~      & 0.062 & 0.062  \\
                        & $\mu$     & ~      & 0.43  & 0.28   \\
\hline
\hline
\end{tabular*}
\end{table}

In the present work, the deformation parameters ($\varepsilon_2$, $\varepsilon_4$)
of $^{166, 168, 170, 172}$Re are taken from Ref.~\cite{Bengtsson1986_ADNDT35-15}
(cf., Table~\ref{tab:def}), which are taken as an average of the adjacent even-even nuclei.
It can be seen that the deformation parameters increase gradually with neutron number.
The Nilsson parameters ($\kappa$, $\mu$) are taken as the
traditional values~\cite{Bengtsson1985_NPA436-14} with a slight change of
neutron $\mu_6$ (modified from 0.34 to 0.28, cf., Table~\ref{tab:ku})
for a better description of the level crossing behavior in Re isotopes.
In addition, the proton orbital $\pi 1/2^-[541]$ is shifted upward
by about 0.7~MeV to avoid the defect caused by the velocity dependent $l^2$ term
in the Nilsson potential at very high-spin region~\cite{Andersson1976_NPA268-205}.
Note that this parameter set is the same as that adopted in our previous work for
$^{166}$Ta~\cite{Zhang2016_SciChinaPMA59-672012}, in which the 2-qp
rotational bands are described quite well.
The proton $N=4, 5$ and neutron $N=5,6$ major shells are adopted to construct the CMPC space.
For both protons and neutrons, the dimensions of the CMPC space are about 1000.
The effective monopole pairing strengths for $^{166, 168, 170, 172}$Re are determined by
fitting the nuclear odd-even mass differences.
In principal, they should be different for each nucleus.
In the present work, the monopole pairing strengths for $^{166, 168, 170, 172}$Re
are chosen as the same value to get a global fit.
They are taken as $G_{\rm p}=0.28$~MeV for protons and $G_{\rm n}=0.48$~MeV
for neutrons, respectively.

\section{Results and discussion}{\label{Sec:Results}}

\begin{figure}[h]
\includegraphics[width=0.8\columnwidth]{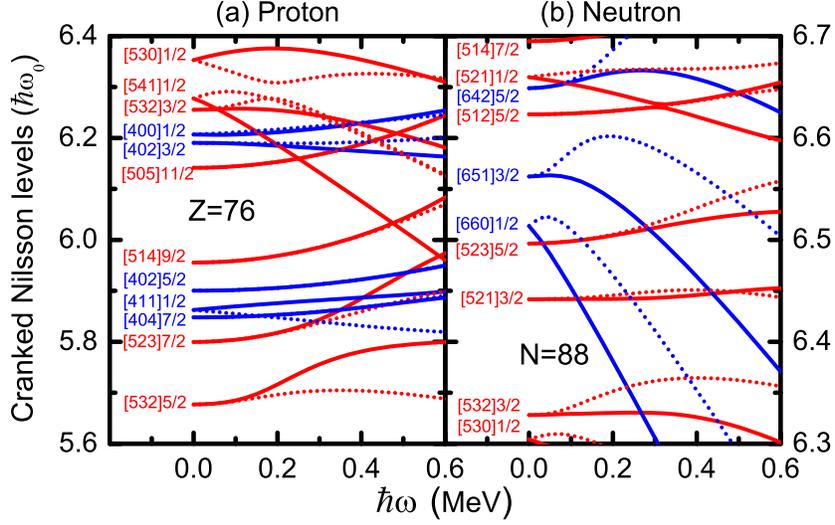}
\centering
\caption{\label{fig1:nil}
The cranked single-particle levels near the Fermi surface of $^{170}$Re
for (a) protons and (b) neutrons.
The positive-parity (negative-parity) levels are displayed by blue (red) lines.
The signature $\alpha=+1/2$ ($\alpha=-1/2$)
levels are displayed by solid (dotted) lines.
}
\end{figure}

Figure~\ref{fig1:nil} shows the proton and neutron cranked Nilsson levels near
the Fermi surface of $^{170}$Re.
The single-particle structures of $^{166, 168, 170, 172}$Re are similar with each other,
so only $^{170}$Re is shown as an example.
It can be seen in Fig.~\ref{fig1:nil} that at low rotational frequencies,
there exist a proton shell gap at $Z=76$ and a neutron shell gap at $N=88$.
The proton and neutron Fermi surfaces of these Re nuclei lie close to the
$\pi h_{11/2}$ and $\nu i_{13/2}$ sub-shells.
$\pi 9/2^-[514]$ ($h_{11/2}$) is occupied as the lowest proton configurations.
This is consistent with the data, which shows that $\pi 9/2^-[514]$
is the yrast bands of odd-$A$ Re isotopes in this mass region.
With the neutron number increasing, the orbitals in neutron $i_{13/2}$ sub-shell
close to the Fermi surface move from $\nu 1/2^+[660]$ to $\nu 5/2^+[642]$.
Therefore, these odd-odd Re isotopes provides a good opportunity to study the dependence of
level crossings on the occupation of specific single-particle orbitals.

\subsection{2-qp rotational bands in $^{166,168,170}$Re}

\begin{figure}[h]
\includegraphics[width=0.8\columnwidth]{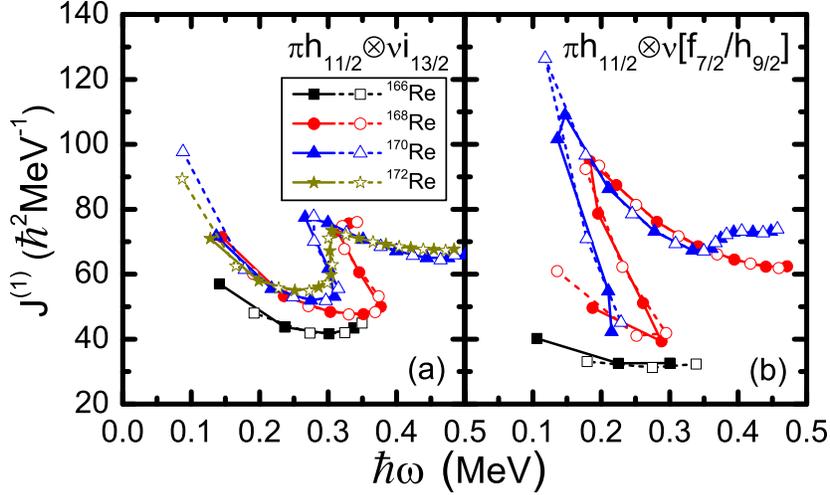}
\centering
\caption{\label{fig2:exp}
The experimental kinematic MOIs for the two configurations (a) $\pi h_{11/2} \otimes \nu i_{13/2}$
and (b) $\pi h_{11/2} \otimes \nu [f_{7/2} / h_{9/2}]$ in
$^{166,168,170, 172}$Re~\cite{Li2015_PRC92-014310, Hartley2016_PRC94-054329,
Hartley2013_PRC87-024315, Hartley2014_PRC90-017301}.
Signature $\alpha=0$ and $\alpha=1$ branches are denoted by solid and open symbols, respectively.
}
\end{figure}

In Ref.~\cite{Li2015_PRC92-014310}, two rotational bands with the
configurations assigned as $\pi h_{11/2} \otimes \nu i_{13/2}$
and $\pi h_{11/2} \otimes \nu [f_{7/2} / h_{9/2}]$ have been observed in $^{166}$Re.
Their bandhead spins are assigned as $8^-$ and $6^+$, respectively.
Similar band structures have also been observed in heavier odd-odd Re
isotopes~\cite{Hartley2016_PRC94-054329, Hartley2013_PRC87-024315, Hartley2014_PRC90-017301}.
Fig.~\ref{fig2:exp} shows the experimental kinematic MOIs for
these two configurations in $^{166, 168, 170, 172}$Re.
It can be seen in Fig.~\ref{fig2:exp}(a) that the experimental MOIs for
$\pi h_{11/2} \otimes \nu i_{13/2}$ in $^{168, 170, 172}$Re are quite similar to each other,
except a larger backbending frequency in $^{168}$Re.
However, the MOIs for this configuration in $^{166}$Re are much smaller.
The same situation happens for $\pi h_{11/2} \otimes \nu [f_{7/2} / h_{9/2}]$ [see Fig.~\ref{fig2:exp}(b)].
In principle, the level structures as well as the MOIs should be similar
for the same configuration in adjacent nuclei.
It also can be seen from Eq.~(\ref{eq:exp-moi}) that for one rotational band,
the extracted kinematic MOIs with rotational frequency are very
sensitive to the bandhead spin.
Therefore, the bandhead spin assignments for these two bands in $^{166}$Re may be questionable.
It should be noted that according to the energy level systematics and the additivity of alignment,
Hartley et al., suggested that the bandhead spins of these two bands observed in $^{166}$Re
should be $10^-$ and $7^+$, respectively~\cite{Hartley2016_PRC94-054329}.

\begin{figure}[h]
\includegraphics[width=0.8\columnwidth]{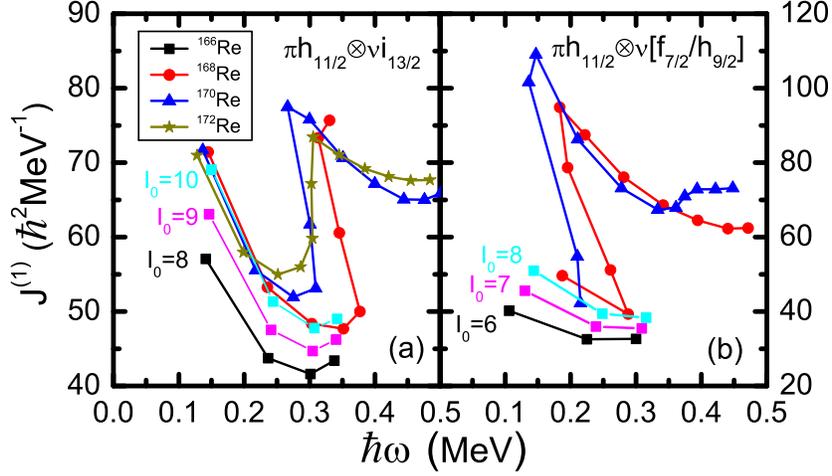}
\centering
\caption{\label{fig3:exp2}
The same as Fig.~\ref{fig2:exp}, but with different bandhead spin
assignments for the two configurations in $^{166}$Re.
Only the branches with signature $\alpha=0$ are shown in this figure.
}
\end{figure}

Figure~\ref{fig3:exp2} shows the extracted MOIs with different bandhead spin
assignments for (a) $\pi h_{11/2} \otimes \nu i_{13/2}$
and (b) $\pi h_{11/2} \otimes \nu [f_{7/2} / h_{9/2}]$ in $^{166}$Re.
The MOIs for the same configuration in  $^{168, 170, 172}$Re are also shown for comparison.
Since there is no signature splitting for these two bands in all these odd-odd Re nuclei,
only the branches with signature $\alpha=0$
are shown in Fig.~\ref{fig3:exp2}.
It can be seen clearly in Figs.~\ref{fig3:exp2}(a) and (b) that the bandhead spins
of these two rotational bands in $^{166}$Re should both be increased by $2\hbar$
to get in consistent with the systematics of the experimental MOIs
for the same configuration in $^{168, 170, 172}$Re.
Therefore, it is reasonable to assign $10^-$ for $\pi h_{11/2} \otimes \nu i_{13/2}$
and $8^+$ for $\pi h_{11/2} \otimes \nu [f_{7/2} / h_{9/2}]$ as the bandhead spins.
Note that the present bandhead spin assignment for $\pi h_{11/2} \otimes \nu i_{13/2}$ is
the same as that suggested by Hartley in Ref.~\cite{Hartley2016_PRC94-054329},
while the bandhead spin for $\pi h_{11/2} \otimes \nu [f_{7/2} / h_{9/2}]$
is increased by 1$\hbar$ compared with Ref.~\cite{Hartley2016_PRC94-054329}.
In the following calculations, the experimental MOIs and alignments
for these two bands in $^{166}$Re will be extracted with bandhead
spins $10^-$ and $8^+$.

\begin{figure}[h]
\includegraphics[width=1.0\textwidth]{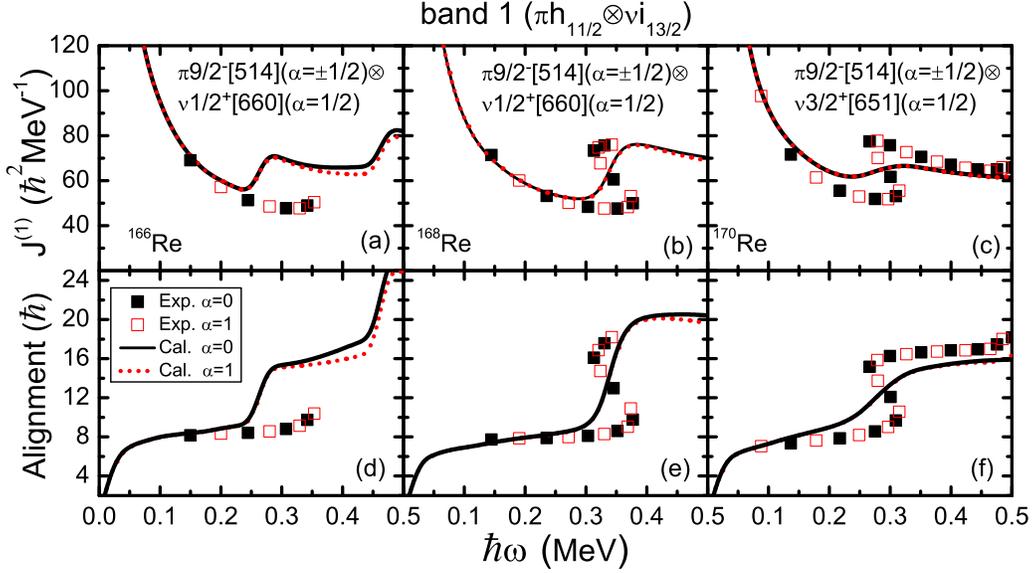}
\centering
\caption{\label{fig4:band1}
Comparison between the experimental and calculated kinematic MOIs $J^{(1)}$
(upper panels) and alignments (lower panels) of the bands with the configuration
$\pi h_{11/2} \otimes \nu i_{13/2}$ (labeled as band 1) in $^{166,168,170}$Re.
The experimental MOIs and alignments are displayed by black solid ($\alpha = 0$)
and red open ($\alpha= 1$) squares, and the calculated values are displayed by
black solid ($\alpha = 0$) and red dotted ($\alpha= 1$) lines.
The alignment $i$ is defined as $i= \langle J_x \rangle-\omega J_0 -\omega^3 J_1$.
The Harris parameters are
$J_{0}=13\ \hbar^{2}\rm{MeV^{-1}}$ and $J_{1}=64\ \hbar^{4}\rm{MeV^{-3}}$
for $^{166}$Re~\cite{Li2015_PRC92-014310},
$J_{0}=17\ \hbar^{2}\rm{MeV^{-1}}$ and $J_{1}=50\ \hbar^{4}\rm{MeV^{-3}}$
for $^{168}$Re~\cite{Hartley2016_PRC94-054329}
and $^{170}$Re~\cite{Hartley2013_PRC87-024315}.
}
\end{figure}

Figure~\ref{fig4:band1} shows the comparison between the experimental and
calculated kinematic MOIs (upper panels) and alignments (lower panels) for
the bands with the configuration $\pi h_{11/2} \otimes \nu i_{13/2}$
(labeled as band 1) in $^{166, 168, 170}$Re.
For odd-odd nucleus, the total signature ($\alpha = 0, 1$) is coupled
by the odd neutron ($\alpha = \pm 1/2$) and odd proton ($\alpha = \pm 1/2$).
It can be seen from the cranked single-particle levels in Fig.~\ref{fig1:nil}(b) that,
all the $\nu i_{13/2}$ orbitals close to the Fermi surface ($\nu 1/2^+ [660]$ and $\nu 3/2^+ [651]$)
have very large signature splitting,
while the experimental data of these bands show nearly no signature splitting.
In addition, Fig.~\ref{fig1:nil}(a) shows that the signature splitting in
$\pi 9/2^-[514]$ $(h_{11/2})$ is very small.
This indicates that band 1 in these three nuclei should be coupled from
$\alpha = 1/2$ (the favored signature) of the odd neutron with
$\alpha =\pm 1/2$ of the odd proton to form the total signature $\alpha = 1, 0$.
Note that although the spherical configurations are the same for these three bands,
their Nilsson configurations are different.
It can be seen in Fig.~\ref{fig1:nil}(b) that the lowest
neutron $\nu i_{13/2}$ orbital is $\nu 1/2^+[660]$ for $N=91$ ($^{166}$Re) and
$N=93$ ($^{168}$Re), but $\nu 3/2^+[651]$ for $N=95$ ($^{170}$Re).
Therefore, the configurations should be
    $\pi 9/2^-[514](\alpha=\pm 1/2) \otimes \nu 1/2^+[660](\alpha=1/2)$ for $^{166, 168}$Re,
and $\pi 9/2^-[514](\alpha=\pm 1/2) \otimes \nu 3/2^+[651](\alpha=1/2)$ for $^{170}$Re.
Note that the bandhead spin of this band in $^{166}$Re
is assigned as $I_{0}=10\hbar$ by Fig.~\ref{fig3:exp2}(a),
which is the same as that suggested by Hartley in Ref.~\cite{Hartley2016_PRC94-054329}.
The experimental data can be reproduced quite well
by the PNC-CSM calculations with the above configurations,
which confirms the configuration and bandhead spin assignments for these bands.
The PNC-CSM calculations predict two sharp upbendings at $\hbar\omega\approx 0.25$~MeV and 0.45~MeV
in $^{166}$Re for this band, which are not observed in experiment.
In addition, the calculated upbending by the PNC-CSM in $^{170}$Re is weaker than the data,
which may be caused by the change of the mean-field after level crossing~\cite{Zhang2020_PRC101-054303}.

\begin{figure}[h]
\includegraphics[width=1.0\textwidth]{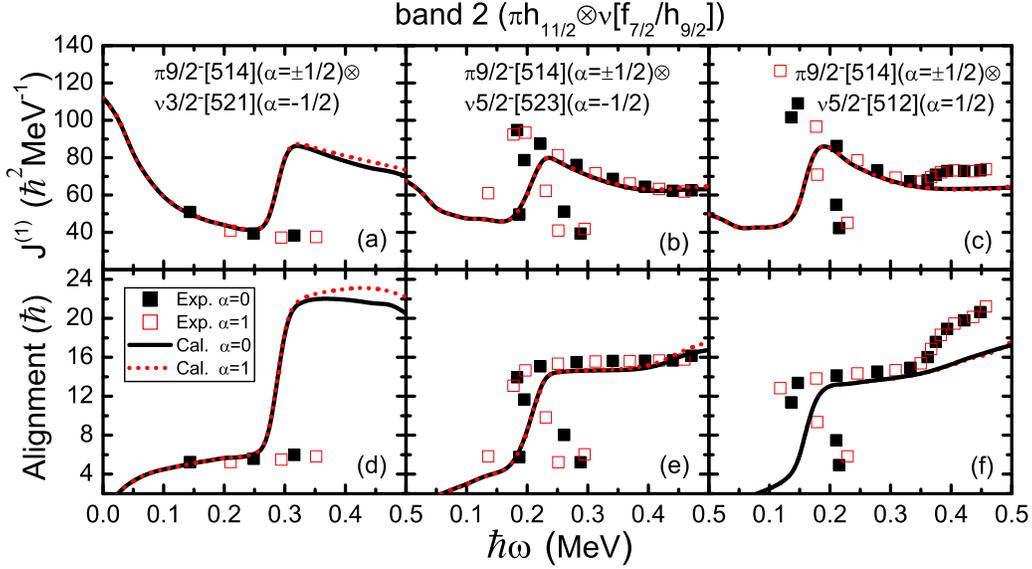}
\centering
\caption{\label{fig5:band2}
The same as Fig \ref{fig4:band1}, but for the bands with the configuration
$\pi h_{11/2} \otimes \nu [f_{7/2} / h_{9/2}]$ (labeled as band 2) in $^{166, 168, 170}$Re.
}
\end{figure}

Close to the neutron Fermi surface, there exist several negative parity
single-particle levels in these Re isotopes, e.g.,
$\nu 3/2^+[521]$ ($h_{9/2}$), $\nu 5/2^+[523]$ ($f_{7/2}$) and
$\nu 5/2^+[512]$ ($h_{9/2}$).
Together with the odd-proton $\pi 9/2^-[514]$ ($h_{11/2}$),
several positive parity 2-qp states can be established.
Fig.~\ref{fig5:band2} is the same as Fig.~\ref{fig4:band1},
but for the bands with the configuration $\pi h_{11/2} \otimes \nu [f_{7/2} / h_{9/2}]$
(labeled as band 2) in $^{166, 168, 170}$Re.
Similar to band 1, there is also nearly no signature splitting in band 2 of these three nuclei.
Fig.~\ref{fig1:nil}(b) shows that signature splitting
exists in all these $\nu [f_{7/2} / h_{9/2}]$ orbitals.
This indicates that band 2 in these three nuclei should be coupled from
the favored signature of the odd neutron with
$\alpha=\pm 1/2$ of the odd proton to form the total signature.
Therefore, the Nilsson configuration
$\pi 9/2^-[514](\alpha=\pm 1/2) \otimes \nu 3/2^+[521](\alpha=-1/2)$ for $^{166}$Re,
$\pi 9/2^-[514](\alpha=\pm 1/2) \otimes \nu 5/2^+[523](\alpha=-1/2)$ for $^{168}$Re,
and $\pi 9/2^-[514](\alpha=\pm 1/2) \otimes \nu 5/2^+[512](\alpha=1/2)$ for $^{170}$Re are assigned.
Note that $f_{7/2}$ and $h_{9/2}$ are pseudo-spin partners,
strong mixing exists in these two orbitals (see the occupation
probabilities in Fig.~\ref{fig7:occu}).
The bandhead spin of this band in $^{166}$Re
is assigned as $I_0=8\hbar$ by Fig.~\ref{fig3:exp2}(b),
which is increased by $1\hbar$ compared with Ref.~\cite{Hartley2016_PRC94-054329}.
It can be seen in Fig.~\ref{fig5:band2} that the experimental kinematic MOIs
and alignments can be well reproduced by the PNC-CSM with the above configurations,
which confirms the configuration and bandhead spin assignments for these bands.
Note that the second upbending observed in $^{170}$Re is not reproduced by the PNC-CSM results.
The calculated rotational frequency of this upbending is about $\hbar\omega\approx$0.52~MeV,
which is much higher than the experimental data.
After comparing the experimental alignments with the adjacent odd-$A$ nuclei $^{167, 169}$Re and $^{167}$W,
the first backbendings in bands 1 and 2 of $^{166, 168, 170}$Re are attributed
to the $BC$ (the alignments of the second and third $i_{13/2}$ quasineutrons)
and $AB$ (alignment of the lowest $i_{13/2}$ quasineutrons) crossing
in Refs~\cite{Hartley2016_PRC94-054329,Hartley2013_PRC87-024315}.
In the following, we will analyze the level crossing mechanism of these bands
in detail using the PNC-CSM.

\begin{figure}[h]
\includegraphics[width=1.0\textwidth]{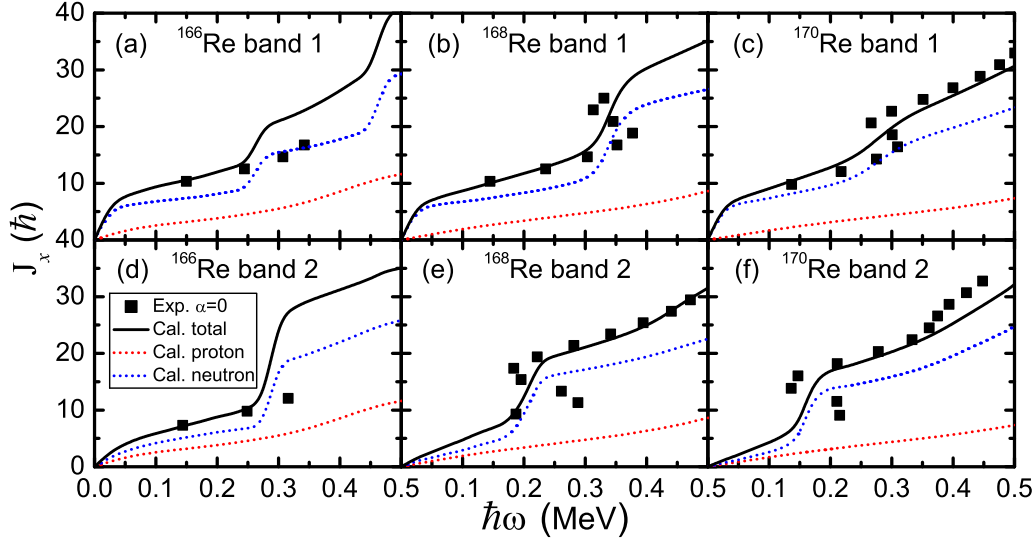}
\centering
\caption{\label{fig6:jx}
The experimental (solid squares) and calculated (black solid lines) angular momentum alignments
$\langle J_x \rangle$ with signature $\alpha=0$
for band 1 ($\pi h_{11/2} \otimes \nu i_{13/2}$, upper panels) and
band 2 ($\pi h_{11/2} \otimes \nu [f_{7/2} / h_{9/2}]$, lower panels) in $^{166, 168, 170}$Re.
Contribution from protons and neutrons is displayed by red and blue dotted lines, respectively.
}
\end{figure}

The experimental and calculated angular momentum alignment
$\langle J_x \rangle$ for band 1 ($\pi h_{11/2} \otimes \nu i_{13/2}$, upper panels) and
band 2 ($\pi h_{11/2} \otimes \nu [f_{7/2} / h_{9/2}]$, lower panels)
in $^{166, 168, 170}$Re with signature $\alpha=0$ are shown in Fig.~\ref{fig6:jx}.
Since there is nearly no signature splitting in all these bands,
we only take $\alpha=0$ branch as an example.
Note that smoothly increasing part of the alignment represented by the
Harris formula ($\omega J_0+\omega^3 J_1$) is not subtracted in this figure.
It can be seen clearly in Fig.~\ref{fig6:jx} that
all the contributions to the backbendings/upbendings in these bands
observed in $^{166, 168, 170}$Re come from the neutrons.
The protons only provide a gradual increase of the angular momentum alignment.
Therefore, in the following, only the neutron part will be discussed.

\begin{figure}[!]
\includegraphics[width=1.0\textwidth]{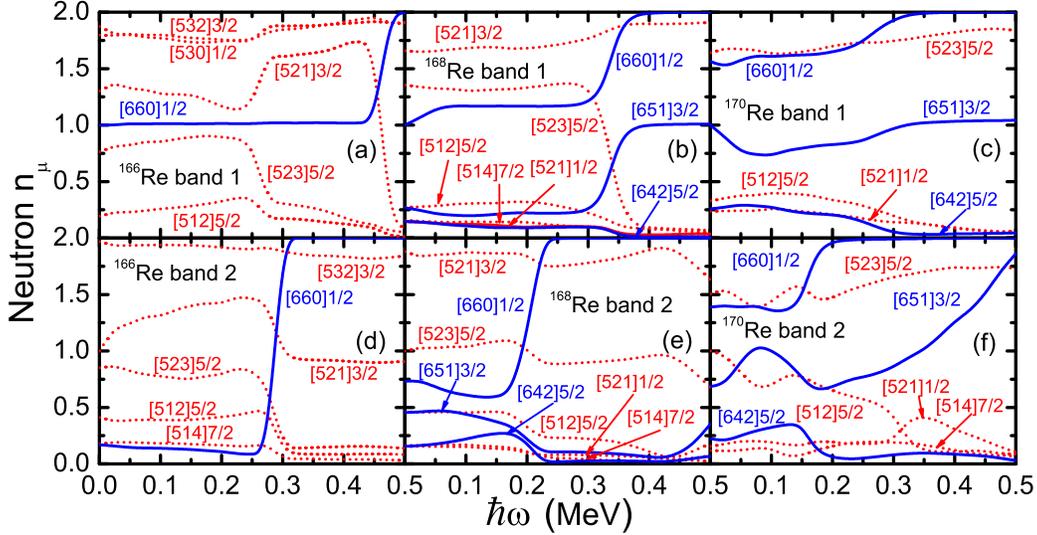}
\centering
\caption{\label{fig7:occu}
The occupation probabilities $n_\mu$ of orbital $\mu$
(including both signature $\alpha=\pm 1/2$) close to the neutron Fermi surface
for band 1 ($\pi h_{11/2} \otimes \nu i_{13/2}$, upper panels) and
band 2 ($\pi h_{11/2} \otimes \nu [f_{7/2} / h_{9/2}]$, lower panels) in $^{166, 168, 170}$Re.
The positive (negative) parity levels are shown by blue solid (red dotted) lines.
}
\end{figure}

Figure~\ref{fig7:occu} shows the occupation probabilities $n_\mu$ of each
orbital $\mu$ close to the neutron Fermi surface for
band 1 ($\pi h_{11/2} \otimes \nu i_{13/2}$, upper panels) and
band 2 ($\pi h_{11/2} \otimes \nu [f_{7/2} / h_{9/2}]$, lower panels) in $^{166,168,170}$Re.
In the PNC-CSM calculations, the particle-number is exactly conversed,
whereas the occupation probabilities of the single-particle orbitals
change with the rotational frequency.
By analyzing the variation of the occupation probabilities with rotational frequency,
we can learn the level crossing mechanism more deeply.
It can be seen in Fig.~\ref{fig7:occu}(a) that
the $n_\mu$ of $\nu 3/2^-[521]$ increases from 1.1 to 1.6
with the rotational frequency increasing from about 0.20~MeV to 0.30~MeV,
while the $n_\mu$ of $\nu 5/2^-[523]$ decreases from 0.9 to 0.3.
At the same time, the $n_\mu$ of some other orbitals
in $N=5$ shell, e.g., $\nu5/2^-[512]$, slightly increase or decrease.
This can be easily understood from the cranked single-particle levels in Fig.~\ref{fig1:nil}(b).
The $\nu 5/2^-[523]$ is slightly above the neutron Fermi surface,
and is partly occupied due to the pairing correlations.
With increasing rotational frequency, this orbital leaves further from the Fermi surface.
Therefore, the occupation probability of this orbital becomes smaller
after the level crossing frequency.
Meanwhile, the occupation probability of $\nu 3/2^-[521]$, which approaches near
to the Fermi surface, becomes larger with increasing rotational frequency.
So the predicted sharp upbending at $\hbar\omega\approx 0.25$~MeV in band 1 of $^{166}$Re
may come from the level crossing of these two pseudo-spin partners.
Around $\hbar\omega\approx 0.45$~MeV, the $n_\mu$ of $\nu 1/2^+[660]$ and $\nu 3/2^+[651]$
increases sharply from about 1.0 to 2.0 and 0.0 to 1.0, respectively, while the
the $n_\mu$ of $\nu 3/2^-[521]$ decreases from 1.7 to 0.1.
This indicates that the second sharp upbending in this band may mainly due to
$\nu 1/2^+[660]$ and $\nu 3/2^+[651]$.
It can be seen in Fig.~\ref{fig7:occu}(b) that the $n_\mu$ of $\nu 1/2^+[660]$ and $\nu 3/2^+[651]$
increases sharply from about 1.2 to 2.0 and 0.2 to 1.0, respectively, while the
the $n_\mu$ of $\nu3/2^-[521]$ decreases from 1.4 to 0.1 around $\hbar\omega\approx 0.35$~MeV.
This indicates that the sharp upbending in band 1 of $^{168}$Re may due to
$\nu 1/2^+[660]$ and $\nu 3/2^+[651]$.
With neutron number increasing, the $\nu 3/2^-[651]$ becomes
the lowest $\nu i_{13/2}$ orbital in $^{170}$Re.
It can be seen in Fig.~\ref{fig7:occu}(c) that the $\nu 3/2^-[651]$
has a strong Coriolis mixing with $\nu 1/2^+[660]$ in band 1 of $^{168}$Re,
and they have a gradual change in the upbending region together with some
orbitals in $N=5$ shell, e.g., $\nu 5/2^-[512]$ and $\nu 5/2^-[523]$.
This indicates that a gradual upbending in band 1 of $^{170}$Re
may mainly due to these $i_{13/2}$ neutrons.

As for the band 2 in $^{166}$Re, the pseudo-spin partners $\nu 3/2^-[521]$ ($h_{9/2}$)
and $\nu 5/2^-[523]$ ($f_{7/2}$)  have a strong mixing [see Fig.~\ref{fig7:occu}(d)].
The $n_\mu$ of $\nu 1/2^+[660]$ increases sharply from nearly zero to 2.0 at
$\hbar\omega\approx 0.25$~MeV, while the $n_\mu$ of $\nu 3/2^-[521]$, $\nu5/2^-[523]$
and $\nu5/2^{-} [512]$ have a sharp decrease.
This indicates that the sharp upbending this band may mainly due to $\nu 1/2^+[660]$.
With neutron number increases by 2, the $\nu 3/2^-[651]$ and $\nu 5/2^-[642]$
get closer to the Fermi surface and are partly occupied in band 2 of $^{168}$Re.
Around $\hbar\omega\approx 0.2$~MeV, the $n_\mu$ of $\nu 1/2^+[660]$ increases
sharply from 0.6 to 2.0, while the $n_\mu$ of $\nu 3/2^-[651]$, $\nu 5/2^-[642]$ and some
orbitals in $N=5$ shell decrease.
This indicates that the sharp upbending in this band may mainly due to these three
$i_{13/2}$ neutrons, i.e., $\nu 1/2^+[660]$, $\nu3/2^-[651]$ and $\nu5/2^-[642]$.
Similar to the band 2 in $^{166}$Re, the pseudo-spin partners $\nu 5/2^-[512]$ ($h_{9/2}$)
and $\nu 5/2^-[523]$ ($f_{7/2}$) also have a strong mixing in this band of $^{170}$Re.
It can be seen in this figure that the the sharp upbending this band may also be
caused by these three $i_{13/2}$ neutrons.

\begin{figure}[h]
\includegraphics[width=1.0\textwidth]{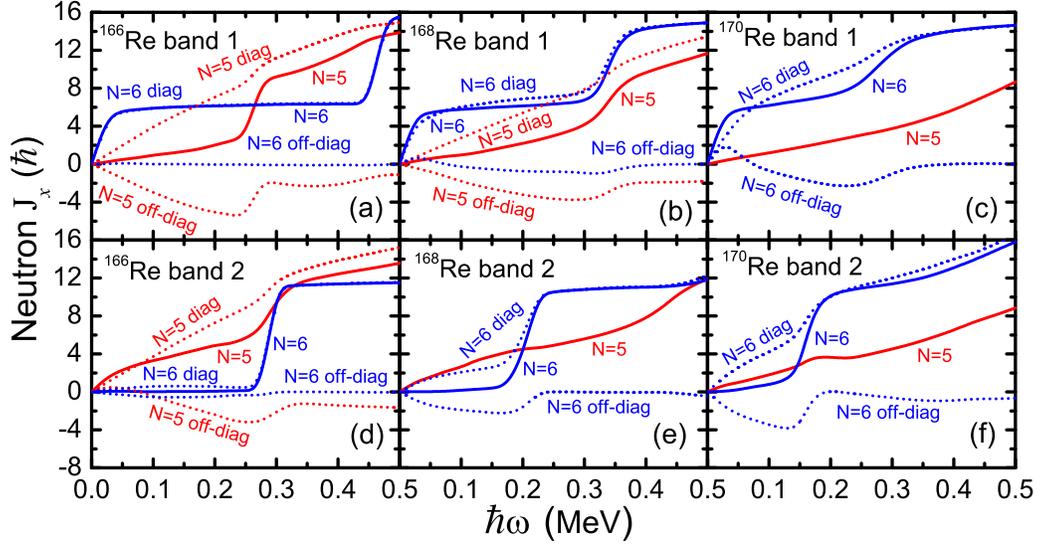}
\centering
\caption{\label{fig8:jxshell}
The contributions of neutron $N = 5, 6$ major shells to the angular momentum alignment
$\langle J_{x} \rangle$ of the band 1 ($\pi h_{11/2} \otimes \nu i_{13/2}$, upper panels) and
band 2 ($\pi h_{11/2} \otimes \nu [f_{7/2} / h_{9/2}]$, lower panels) in $^{166,168,170}$Re.
Red and blue solid lines are used for the $N = 5$ and $N = 6$ shells, respectively.
The contributions of diagonal [$\sum_{\mu} j_x(\mu)$] and off-diagonal
[$\sum_{\mu<\nu} j_x(\mu\nu)$] parts in Eq.~(\ref{eq:jx}) are shown by dotted lines.
}
\end{figure}

To have a clearer understanding of the level crossing mechanism in these rotational bands,
the contribution of neutron $N = 5, 6$ major shells to the angular momentum alignment
$\langle J_x \rangle$ of band 1 ($\pi h_{11/2} \otimes \nu i_{13/2}$, upper panels)
and band 2 ($\pi h_{11/2} \otimes \nu [f_{7/2} / h_{9/2}]$, lower panels)
in $^{166,168,170}$Re are shown in Fig.~\ref{fig8:jxshell}.
It can be seen in Fig.~\ref{fig8:jxshell}(a) that the first upbending in band 1 of $^{166}$Re
is caused by the neutron $N=5$ shell, and the off-diagonal part contributes to this upbending.
This is due to the rearrangement of neutron occupations in these pseudo-spin partners.
The second upbending is mainly due to the diagonal part of $N=6$ major shell.
Fig.~\ref{fig8:jxshell}(b) shows that both the $N=5$ and 6 major shells
contribute to the upbending in band 1 of $^{168}$Re.
The main contribution comes from the off-diagonal part of $N=5$
and the diagonal part of $N=6$ major shells.
For the upbending in band 1 of $^{168}$Re, the main contribution comes from
the off-diagonal part of $N=6$ major shell.
It can be seen that the contribution from $N=5$ major shell decreases with
increasing neutron number.
This is because the occupation of the low-$\Omega$ orbital $\nu 3/2^-[521]$ ($h_{9/2}$),
which contributes a remarkable amount of angular momentum,
is getting larger and larger with increasing neutron number.
This is different with the traditional point of view that only the $\nu i_{13/2}$
orbitals contribute to the first backbending/upbending in the rare-earth nuclei.
Similar to band 1, the contribution from $N=5$ major shell also decreases with
increasing neutron number in band 2 of $^{166,168,170}$Re [see Figs.~\ref{fig8:jxshell}(d), (e), (f)].
The neutron $i_{13/2}$ orbital is not blocked in this configuration,
so the level crossing mechanism should be different from that in band 1.
It can be see in Figs.~\ref{fig8:jxshell}(d), (e), (f) that except $^{166}$Re,
both the diagonal and the off-diagonal parts of $N=6$ shell contribute to
the upbending in band 2 of $^{168,170}$Re.
It can be seen that with neutron number increasing,
the off-diagonal contribution from $N=6$ shell becomes larger and larger,
while the diagonal contribution becomes  smaller and smaller.
This is because the neutron Fermi surface approaches the middle of the
$\nu i_{13/2}$ sub-shell with increasing neutron number.

\begin{figure}[!]
\includegraphics[width=1.0\textwidth]{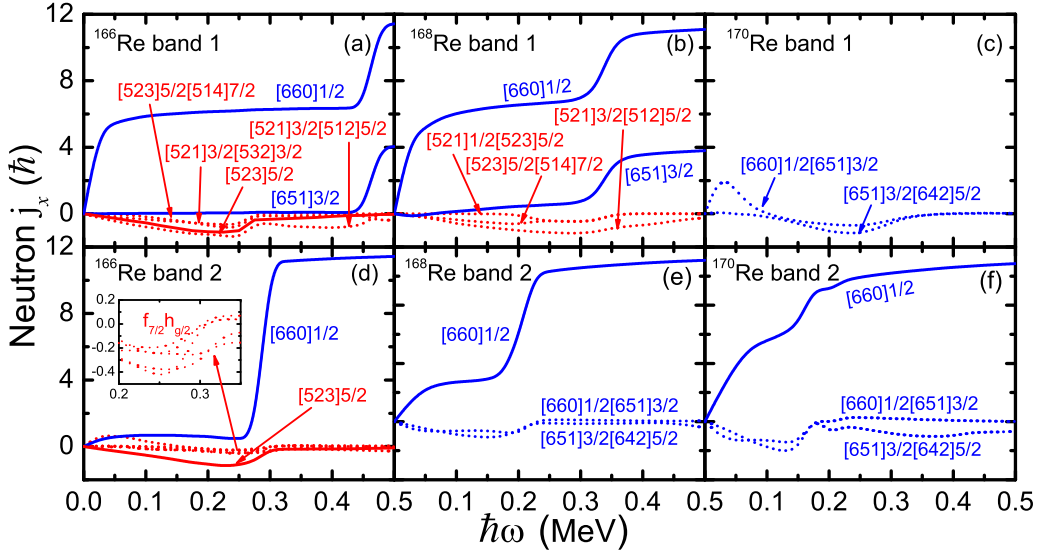}
\centering
\caption{\label{fig9:jxborb}
Contribution of each neutron orbital in the
$N = 5$ and $N=6$ major shell for (a) band 1 and (b) band 2
to the angular momentum alignments $\langle J_x \rangle$ in $^{166, 168, 170}$Re.
The diagonal [$j_x(\mu)$] and off-diagonal [$j_x(\mu\nu)$] parts
in Eq.~(\ref{eq:jxorb}) are denoted by solid and dotted lines, respectively.
}
\end{figure}

By analyzing the contribution of each diagonal [$j_x(\mu)$]
and off-diagonal [$j_x(\mu\nu)$] parts in Eq.~(\ref{eq:jxorb})
in bands 1 and 2 of $^{166, 168, 170}$Re,
the level crossing mechanism can be understood thoroughly.
Fig.~\ref{fig9:jxborb} shows the contribution of each neutron orbital in the
$N = 5$ and $N=6$ major shell for band 1 (upper panels) and band 2 (lower panels)
to the angular momentum alignments $\langle J_x \rangle$ in $^{166, 168, 170}$Re.
It can be seen in Fig.~\ref{fig9:jxborb}(a) that for band 1 in $^{166}$Re,
the contribution from $N=5$ shell to the alignment gain after level crossing
comes from the off-diagonal parts $j_x(3/2^-[521]5/2^-[512])$,
$j_x(3/2^-[521]3/2^-[532])$, $j_x(5/2^-[523]7/2^-[514])$,
and the diagonal part $j_x(5/2^-[523])$.
They all originate from the pseudo-spin partners $h_{9/2}$ and $f_{7/2}$.
The diagonal parts $j_x(1/2^+[660])$ and $j_x(3/2^+[651])$ in $N=6$ shell
contribute to the second upbending.
For band 1 in $^{168}$Re [Fig.~\ref{fig9:jxborb}(b)],
the contribution from $N=5$ shell to the alignment gain after level crossing
comes from the off-diagonal parts $j_x(3/2^-[521]5/2^-[512])$,
$j_x(5/2^-[523]7/2^-[514])$, and $j_x(1/2^-[521]5/2^-[523])$.
They all come from the pseudo-spin partners $h_{9/2}$ and $f_{7/2}$, except $1/2^-[521]$ ($p_{3/2}$).
The main contribution from $N=6$ shell comes from the
diagonal parts $j_x(1/2^+[660])$ and $j_x(3/2^+[651])$.
The off-diagonal part $j_x(3/2^+[651]5/2^+[642])$ contributes a little.
As for the band 1 in $^{170}$Re [Fig.~\ref{fig9:jxborb}(c)],
only the off-diagonal parts $j_x(1/2^+[660]3/2^+[651])$ and $j_x(3/2^+[651]5/2^+[642])$ contribute
to the upbending.

It can be seen in Fig.~\ref{fig9:jxborb}(e) that the alignment gain after level
crossing in band 2 of $^{166}$Re mainly comes from the diagonal part $j_x(1/2^+[660])$.
The interference terms between several orbitals from the pseudo-spin
partners $h_{9/2}$ and $f_{7/2}$ contribute a lot to the upbending,
although each of them only contribute a little [see the inset of Fig.~\ref{fig9:jxborb}(e)].
The diagonal part $j_x(5/2^-[523])$ also has a little contribution.
For band 2 in $^{168}$Re [Fig.~\ref{fig9:jxborb}(e)], the diagonal part $j_x(1/2^+[660])$
contributes a lot to the upbending.
The off-diagonal parts $j_x(1/2^+[660]3/2^+[651])$ and $j_x(3/2^+[651]5/2^+[642])$
also have remarkable contribution.
Band 2 in $^{170}$Re is similar to that in $^{168}$Re.
The off-diagonal parts $j_x(1/2^+[660]3/2^+[651])$, $j_x(3/2^+[651]5/2^+[642])$ and
the diagonal part $j_x(1/2^+[660])$ contribute to the upbending.
Therefore, the level crossing mechanism in these bands is understood clearly.

\subsection{2- and 4-qp rotational bands in $^{172}$Re}

\begin{table}[h]
 \centering
 \caption{\label{tab:172config}
The configurations of the five rotational bands observed in $^{172}$Re
proposed in Ref.~\cite{Hartley2014_PRC90-017301}.}
\begin{tabular*}{0.8\columnwidth}{c@{\extracolsep{\fill}}ll}
\hline
\hline
Band   & Configuration\\
\hline
band 1 & $\pi h_{9/2} (\alpha=1/2) \otimes \nu i_{13/2} (\alpha=\pm 1/2)$          \\
band 2 & $\pi h_{11/2}(\alpha=\pm1/2) \otimes \nu i_{13/2} (\alpha=1/2)$          \\
band 3 & $\pi h_{9/2} (\alpha=1/2) \otimes \nu [f_{7/2}/h_{9/2}](\alpha=\pm 1/2)$ \\
band 4 & $\pi h_{9/2}  \otimes \nu^3 (p_{3/2}AB)$ (favored signature)   \\
band 5 & $\pi h_{9/2}  \otimes \nu^3 (p_{3/2}AB)$ (unfavored signature) \\
\hline
\hline
\end{tabular*}
\end{table}

With deformation increasing, the $\pi 1/2^{-} [541]$ ($h_{9/2}$) gets closer
to the proton Fermi surface in $^{172}$Re and several 2- and 4-qp bands based on
this quasi-proton have been observed experimentally~\cite{Hartley2014_PRC90-017301}.
According to the alignment properties and observed band crossings,
proper spherical configurations are proposed for these multi-qp
rotational bands by Hartley et al.~\cite{Hartley2014_PRC90-017301},
which are shown at Table~\ref{tab:172config}.
Bands 1, 2 and 3 are 2-qp structures, and bands 4 and 5 are 4-qp structures.
Note that band 4 is observed following the $AB$ crossing and the
configuration $\pi h_{9/2} \otimes \nu^3 (p_{3/2}AB)$ is tentatively assigned
due to the insufficient spectroscopic information.
In addition, the unfavored signature of $\pi h_{9/2} \otimes \nu^3 (p_{3/2}AB)$
would have higher energy compared to the favored one,
and both $\pi h_{9/2}$ and $\nu p_{3/2}$ have significant signature splitting.
So Ref.~\cite{Hartley2014_PRC90-017301} did not assign firmly this configuration for band 5.
Later on, we will check whether the configurations assigned for these bands
are reasonable by comparing their experimental and calculated MOIs and alignments.

\begin{figure}[h]
\includegraphics[width=0.6\columnwidth]{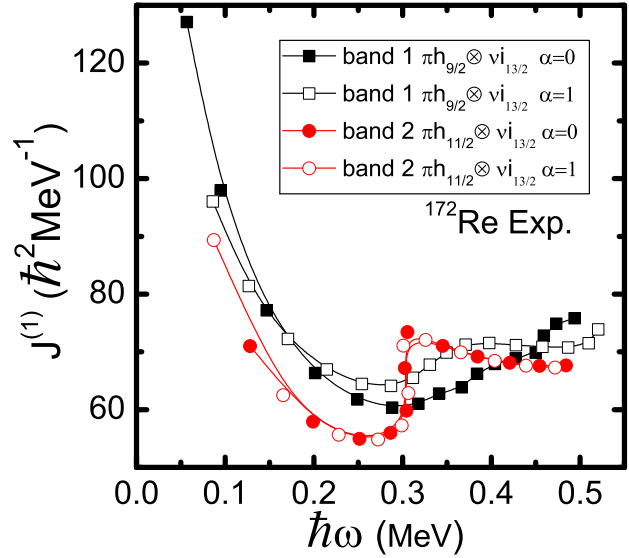}
\centering
\caption{\label{fig10:172exp}
Experimental MOIs of band 1 ($\pi h_{9/2} \otimes \nu i_{13/2}$)
and band 2 ($\pi h_{11/2} \otimes \nu i_{13/2}$).
Solid and open symbols denote the signature $\alpha=0$ and $\alpha=1$ branches, respectively.
}
\end{figure}

Figure~\ref{fig10:172exp} shows the experimental MOIs of
band 1 ($\pi h_{9/2} \otimes \nu i_{13/2}$) and
band 2 ($\pi h_{11/2} \otimes \nu i_{13/2}$) in $^{172}$Re.
It is well known that the proton $h_{11/2}$ level crossing
appears at very high-spin region in the rare-earth nuclei,
especially this orbital is blocked in band 2 of $^{172}$Re,
so the level crossings in these two bands should be caused by neutron.
Since the neutron configuration is the same for
band 1 with $\alpha=1$ (open squares in Fig.~\ref{fig10:172exp}) and band 2,
their level crossing should appear at similar frequency.
However, it can be seen in Fig.~\ref{fig10:172exp} that the level crossing frequency of
signature $\alpha=1$ branch of band 1 in $^{172}$Re
is about 0.34~MeV, which is larger than that in band 2 (the level crossing
frequency is about 0.30~MeV).
Similar delayed crossing appears at signature $\alpha=0$ branch
of band 1 (solid squares in Fig.~\ref{fig10:172exp}),
which is observed at a higher frequency of 0.38~MeV.
This is also slightly higher than the level crossing frequency at the $\nu i_{13/2}$ band
with unfavored signature in the adjacent $^{171}$W~\cite{Espino1994_NPA567-377}.
It is well known that the $h_{9/2}$ proton has very strong prolate deformation
driving effects and drives the nucleus to a slightly larger
deformation~\cite{Nazarewicz1990_NPA512-61, Jensen1991_ZPA340-351, Warburton1995_NPA591-323, Jensen2001_NPA695-3},
which in turn results in this delay in the crossing frequency.
The deformation driving effects of the the $h_{9/2}$ proton make the
level crossings extremely complicated, especially in odd-odd nuclei.
Therefore, in the following calculation, the deformation parameter is chosen as 0.234
when the $\pi 1/2^-[541]$ ($h_{9/2}$) is involved (bands 1, 3-5),
which is increased by 10\% compared with the Lund systematics~\cite{Bengtsson1986_ADNDT35-15}.

\begin{figure}[h]
\includegraphics[width=1.0\textwidth]{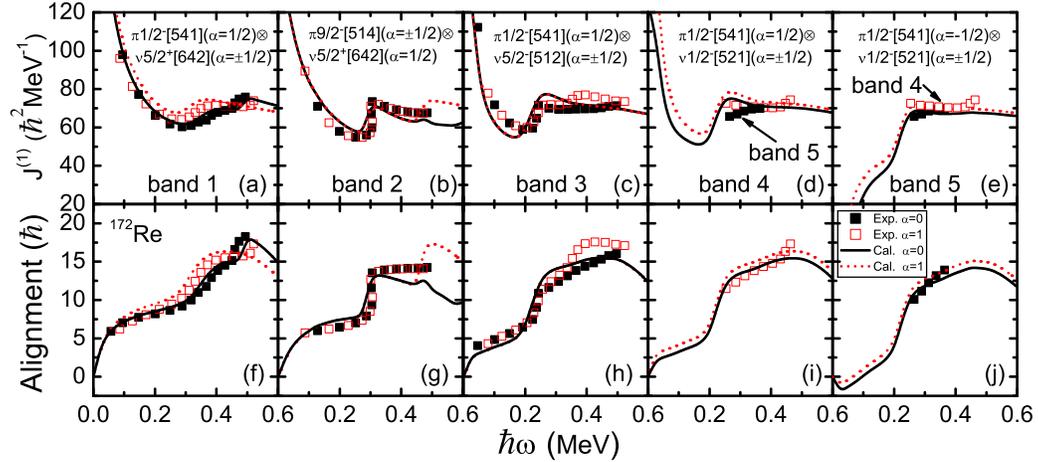}
\centering
\caption{\label{fig11:172moi}
The comparison between the experimental~\cite{Hartley2014_PRC90-017301}
and calculated kinematic MOIs $J^{(1)}$ (upper panels) and alignments (lower panels)
of the five rotational bands observed in $^{172}$Re.
The experimental MOIs and alignments are displayed by black solid and red open squares
for $\alpha=0$ and $\alpha=1$, respectively.
Corresponding calculated values are displayed by black solid and red dotted lines.
The alignments $i=\langle J_x \rangle -\omega J_0 -\omega^3 J_1$
and the Harris parameters are $J_{0} =23~\hbar^2{\rm MeV^{-1}}$ and
$J_1=65~\hbar^4{\rm MeV^{-3}}$~\cite{Hartley2014_PRC90-017301}.
When the $\pi 1/2^-[541]$ is involved (bands 1, 3-5),
the deformation parameter is chosen as 0.234, which is increased by 10\% compared
with the Lund systematics~\cite{Bengtsson1986_ADNDT35-15}.
}
\end{figure}

Figure~\ref{fig11:172moi} shows the comparison between the experimental
and calculated kinematic MOIs $J^{(1)}$ (upper panels) and alignments (lower panels) of
the five rotational bands observed in $^{172}$Re.
The corresponding Nilsson configurations are also assigned for these bands.
It can be seen that the experimental MOIs and alignments,
as well as the level crossings can be well reproduced by the PNC-CSM,
which in turn confirms their configuration assignments.
The results also indicate that a 10\% increase of the deformation is appropriate
to describe the bands with $\pi 1/2^-[541]$ involved.
For band 1, the experimental MOIs exhibit an obvious signature splitting,
especially at high-spin region.
Note that both the proton $\pi 1/2^-[541]$ and the neutron $\nu 5/2^+[642]$
orbitals have significant signature splitting (see Fig.~\ref{fig1:nil}).
However, according to the alignment behaviour of $\pi 1/2^-[541]$
and the level crossing behaviour of these two signature branches,
it is easily to assign this band as the favored signature of
$\pi 1/2^-[541] (\alpha=1/2)$ coupled with the $\alpha=\pm 1/2$ of $\nu 5/2^+[642]$.
It can be seen that for the $\alpha=0$ sequence,
both two upbendings are reproduced quite well by the PNC-CSM,
while the second upbending in the $\alpha=1$ sequence appears at a much higher
frequency in the PNC-CSM calculations compared with the experimental data.
For band 2, the experimental MOIs exhibit nearly no signature splitting.
The proton $\pi 9/2^-[514]$ orbital only shows a very small splitting
at high-spin region [see Fig.~\ref{fig1:nil}(a)].
So this band is assigned as the favored signature of
$\nu5/2^+[642] (\alpha=1/2)$ coupled with the $\alpha=\pm 1/2$ of $\pi 9/2^-[514]$.
It can be seen that the PNC-CSM predicts a second upbending a the
$\alpha=1$ sequence in this band, which is not observed by the experiments.
This upbending is caused by the $h_{9/2}$ proton, i.e., the level crossing
between the $\alpha=1/2$ of $\pi 9/2^-[514]$ and $\pi 1/2^-[541]$.
The experimental MOIs of band 3 only exhibit a signature splitting
after the first level crossing.
Similar to band 1, the configuration of this band is assigned as
the favored signature of $\pi 1/2^-[541] (\alpha=1/2)$ coupled with
the $\alpha=\pm 1/2$ of $\nu5/2^-[512]$.
Note that $\nu5/2^-[512]$ has very small signature splitting [see Fig.~\ref{fig1:nil}(b)],
so both two signature sequences in the PNC-CSM calculations are nearly degenerate.
Therefore, the second upbending in the $\alpha=1$ sequence in this band is not
reproduced by the PNC-CSM.
This needs further investigation.

Experimentally, only one signature sequence is observed
both in bands 4 ($\alpha=0$) and 5 ($\alpha=1$).
D. J. Hartley et al., suggested that these two bands my be
based on the favored (band 4) and unfavored (band 5) signature
sequence of $\pi h_{9/2} \otimes \nu^3 (p_{3/2}AB)$.
Since both $\pi 1/2^-[541]$ and $\nu 1/2^-[521]$ $(p_{3/2})$ orbitals have signature splitting,
four rotational bands with different MOIs can be established.
It can be seen that close to the bandhead region,
the experimental MOIs of band 4 and band 5 have totally different behavior.
One decreases and another increases with rotational frequency.
In Fig.~\ref{fig11:172moi}(d) the PNC-CSM calculations with the
configuration $\pi 1/2^-[541](\alpha=1/2) \otimes \nu 1/2^-[521](\alpha=\pm1/2)$
are compared with the data.
It can be seen that after level crossing, the configuration
$\pi 1/2^-[541](\alpha=1/2)\otimes \nu 1/2^-[521](\alpha=1/2)$
can reproduce the band 4 quite well, while the calculated results with the configuration
$\pi 1/2^-[541](\alpha=1/2)\otimes \nu 1/2^-[521](\alpha=-1/2)$ disagree with band 5.
This indicates that the configuration of band 4 may be
$\pi 1/2^-[541](\alpha=1/2)\otimes \nu^3 1/2^-[521]AB(\alpha=1/2)$.
Furthermore, the PNC-CSM calculations with the configuration
$\pi 1/2^-[541](\alpha=-1/2) \otimes \nu 1/2^-[521](\alpha=\pm1/2)$
are compared with the data in Fig.~\ref{fig11:172moi}(e).
It can be seen that after level crossing, the configuration
$\pi 1/2^-[541](\alpha=-1/2)\otimes \nu 1/2^-[521](\alpha=1/2)$
can reproduce the band 5 quite well.
This indicates that the configuration of band 5 may be
$\pi 1/2^-[541](\alpha=-1/2)\otimes \nu^3 1/2^-[521]AB(\alpha=1/2)$.
It seems that band 4 can also be reproduced by the configuration
$\pi 1/2^-[541](\alpha=-1/2)\otimes \nu 1/2^-[521](\alpha=-1/2)$.
However, both $\pi 1/2^-[541](\alpha=-1/2)$ and $\nu 1/2^-[521](\alpha=-1/2)$ are
unfavored signature branches, the energy of this coupling mode should be much higher
than the coupling mode $\pi 1/2^-[541](\alpha=1/2) \otimes \nu 1/2^-[521](\alpha=1/2)$
with both favored signature.
Therefore, we tentatively assign the configuration of band 4 as
$\pi 1/2^-[541](\alpha=1/2)\otimes \nu^3 1/2^-[521]AB(\alpha=1/2)$.
Note that the two signature branches of $\pi 1/2^-[541]$
have quite different behavior at low-spin region.
Therefore, the coupling mode of band 4 can be determined firmly if we have more
spectroscopic information about the low-spin region for this band.

\begin{figure}[h]
\includegraphics[width=1.0\textwidth]{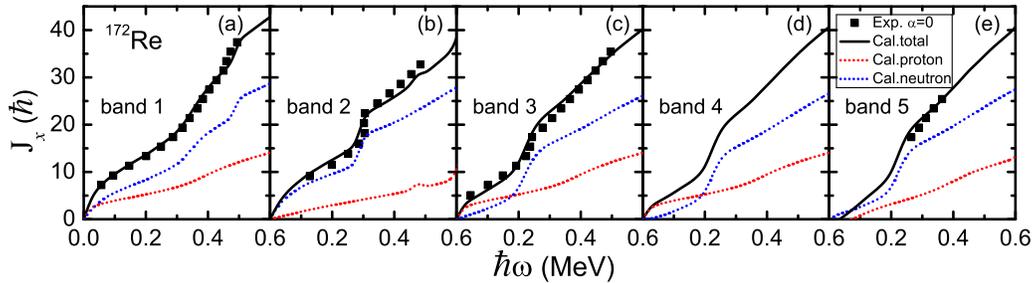}
\centering
\caption{\label{fig12:172jx}
The experimental (solid squares) and calculated (black solid line) angular
momentum alignments $\langle J_x \rangle$ for bands 1-5 with $\alpha=0$ in $^{172}$Re.
Proton and neutron contributions to $\langle J_x \rangle$
are shown by red and blue dotted lines, respectively.
}
\end{figure}

Figure~\ref{fig12:172jx} shows the experimental (solid squares) and calculated
(black solid line) angular momentum alignments $\langle J_x \rangle$ for bands 1-5 in $^{172}$Re.
Here we take the $\alpha=0$ branch in each band as an example.
It can be seen clearly in Fig.~\ref{fig12:172jx} that
all the contributions to the backbendings/upbendings in these five bands come from the neutrons.
The protons only provide a gradual increase of the angular momentum alignment.
Therefore, in the following only the neutron part will be discussed.

\begin{figure}[h]
\includegraphics[width=1.0\textwidth]{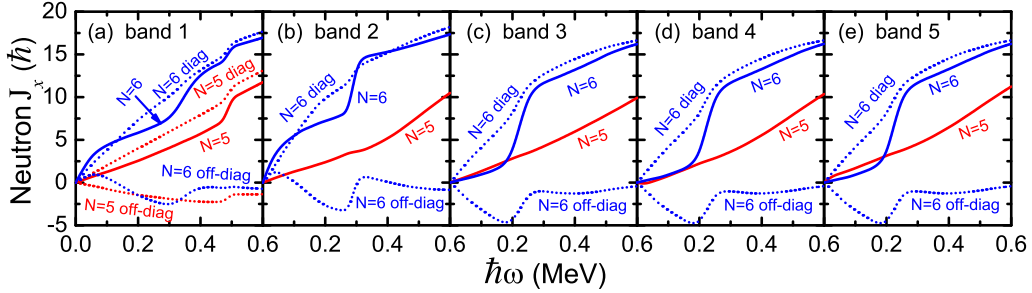}
\centering
\caption{\label{fig13:172jxshell}
The contributions of neutron $N = 5, 6$ major shells to the angular momentum alignment
$\langle J_x \rangle$ for bands 1-5 with $\alpha=0$ in $^{172}$Re.
Red and blue solid lines are used for the $N = 5$ and $N = 6$ shells, respectively.
The contributions of diagonal [$\sum_{\mu} j_x(\mu)$] and off-diagonal
[$\sum_{\mu<\nu} j_x(\mu\nu)$] parts are shown by dotted lines.
}
\end{figure}

The contributions of neutron $N = 5, 6$ major shells to the angular momentum alignment
$\langle J_x \rangle$ for bands 1-5 in $^{172}$Re are shown in Fig.~\ref{fig13:172jxshell}.
It can be seen that the $N=5$ shell has no contribution to the backbendings/upbendings
in these bands except the second upbending in band 1 [see Fig.~\ref{fig13:172jxshell}(a)],
in which both the diagonal and off-diagonal parts have similar contribution.
Note that the diagonal part of $N=6$ shell also contributes a little to the second upbending in band 1.
It also can be seen that no matter the $\nu 5/2^+[642]$ ($i_{13/2}$) being blocked or not,
the diagonal part of $N=6$ shell only contributes a little to the first backbendings/upbendings.
The main contribution comes from the off-diagonal part of $N=6$ shell.

\begin{figure}[h]
\includegraphics[width=1.0\textwidth]{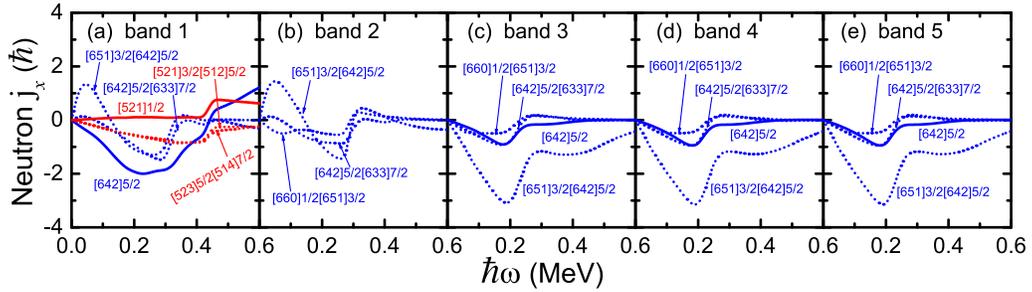}
\centering
\caption{\label{fig14:172jxorb}
Contribution of each neutron orbital in the
$N = 5$ and $N=6$ major shell for bands 1-5 with $\alpha=0$ to the angular momentum
alignments $\langle J_x \rangle$ in $^{172}$Re.
The diagonal [$j_x(\mu)$] and off-diagonal [$j_x(\mu\nu)$] parts
in Eq.~(\ref{eq:jxorb}) are denoted by solid and dotted lines, respectively.
}
\end{figure}

Figure~\ref{fig14:172jxorb} shows the contribution of each neutron orbital in the
$N = 5$ and $N=6$ major shells for bands 1-5 with $\alpha=0$ to the angular momentum
alignments $\langle J_x \rangle$ in $^{172}$Re.
It can be seen in Fig.~\ref{fig14:172jxorb}(a) that the off-diagonal parts
$j_x(3/2^+[651]5/2^+[642])$ and $j_x(5/2^+[642]7/2^+[633])$ contribute
to the first upbending in band 1.
The main contribution to the second upbending comes from the $N=5$ major shell.
The diagonal part $j_x(1/2^-[521])$ and the off-diagonal parts
$j_x(3/2^-[521]5/2^-[512])$  and $j_x(5/2^-[523]7/2^-[514])$
contribute to the second upbending.
The diagonal part $j_x(5/2^+[642])$ in $N=6$ major shell also have contribution.
For band 2, the the off-diagonal parts $j_x(1/2^+[660]3/2^+[651])$,
$j_x(3/2^+[651]5/2^+[642])$, and $j_x(5/2^+[642]7/2^+[633])$
contribute to the upbending.
In bands 3-5, the neutron orbital $5/2^+[642]$ ($\nu i_{13/2}$) is not blocked.
It can be seen in Figs.~\ref{fig14:172jxorb}(c), (d) and (e) that the contribution
of each neutron orbital to $\langle J_x \rangle$ in these three bands are quite similar with each other.
The main contribution comes from the off-diagonal part $j_x(3/2^+[651]5/2^+[642])$.
Moreover, the off-diagonal parts $j_x(1/2^+[660]3/2^+[651])$, and $j_x(5/2^+[642]7/2^+[633])$,
and the diagonal part $j_x(5/2^+[642])$ also have remarkable contributions.

\section{Summary}{\label{Sec:Summary}}

In summary, the recently observed two and four-quasiparticle high-spin
rotational bands in the odd-odd nuclei $^{166,168,170,172}$Re are investigated
using the cranked shell model with pairing correlations treated
by a particle-number conserving method, in which the particle-number is
strictly conserved and the Pauli blocking effects are taken into account exactly.
The Nilsson configurations for these multi-quasiparticle bands have been assigned.
The experimental moments of inertia and alignments of these bands can be well reproduced
by the present calculation if appropriate bandhead spins and configurations are assigned,
which in turn confirms the spin and configuration assignments.
It is found that the bandhead spins of those two rotational
bands observed in $^{166}$Re with the configurations assigned as
$\pi h_{11/2} \otimes \nu i_{13/2}$ and $\pi h_{11/2} \otimes \nu [f_{7/2} / h_{9/2}]$
should be assigned as $10^-$ and $8^+$ (both be increased by $2\hbar$ compared with
those given in Ref.~\cite{Li2015_PRC92-014310}) to get in consistent with the
systematics of the experimental and calculated moments of inertia for the
same configuration in $^{168,170,172}$Re.
The variations of the backbendings/upbendings with increasing neutron number
in these nuclei are also investigated.
The level crossing mechanism is well understood by
analysing the variations of the occupation probabilities of the
single-particle states close to the Fermi surface and their
contributions to the angular momentum alignment with rotational frequency.
In addition, the influence of the deformation driving effects of the proton
$\pi 1/2^-[541]$ ($h_{9/2}$) orbtial on the level crossing in $^{172}$Re are also discussed.

\section*{Acknowledgement}

This work is supported by National Natural Science Foundation of
China (No. 11875027, 11775112, 11775026, 11775099, 11975096),
and the Fundamental Research Funds for the Central Universities (2021MS046).


\end{document}